\documentclass[final,1p,times]{elsarticle} 
\usepackage{graphicx}
\usepackage{amssymb} 
\usepackage{amsthm} 
\usepackage{lineno}
\usepackage{amsmath}
\usepackage{float}

\def\be{\begin{equation}}
\def\ee{\end{equation}}
\def\ba{\begin{align}}
\def\ea{\end{align}}

\def\eq#1{{Eq.~(\ref{#1})}}
\def\fig#1{{Fig.~\ref{#1}}}
\def\Fig#1{{Fig.~\ref{#1}}}
\def\llangle{\left\langle}
\def\rrangle{\right\rangle}

\def\drangle{\rangle\!\rangle}
\def\dlangle{\langle\!\langle}

\journal{Nuclear Physics A} 

\begin{document}

\begin{frontmatter} 

% Your Title - please insert
\title{Non-linear response and event plane correlations}

%% Single author (and collaboration) - please insert
\author{Derek Teaney and Li Yan\fnref{au1}}
\fntext[au1] {Speaker, email: li.yan@stonybrook.edu}
\address{Department of Physics and Astronomy, Stony Brook University, Stony Brook, New York, 11790, USA}

%% Multiple authors
%\author[auth2]{Marcus Junius Brutus}
%\address[auth1]{Somewhere, Rome}
%\address[auth2]{Somewhere else, Rome

\begin{abstract} 
We apply a non-linear flow response formalism to the recently measured event plane correlations. 
We find that the observed event plane correlations can be understood as
an average of the linear and quadratic response.
\end{abstract} 

\end{frontmatter} % do not change

%% linenumbers are useful for reviewing process
%\linenumbers

\section{Introduction}

The Quark-Gluon Plasma (QGP), as observed in heavy-ion collisions at RHIC and  the
LHC, exhibits strong collective flow \cite{Arsene:2004fa,Adcox:2004mh,Muller:2012zq,Gyulassy:2004zy}. 
%One of the
%most pressing tasks in the field of heavy-ion collisions  is 
%to determine the shear viscosity of the plasma from 
%this macroscopic response.  
The observed flow pattern is characterized by long range correlations in the final state
particle spectrum. Indeed, ATLAS recently measured event plane correlations \cite{Jia:2012sa}, and significant
two-plane and three-plane correlations were observed. In comparison to the 
$v_n$ measurements, the 
event plane correlations
provide additional  insight into the origin of initial state fluctuations and
additional constraints on the shear viscosity of the QGP. Although these correlations 
can be simulated with event-by-event hydrodynamics \cite{Qiu:2012uy}, 
it is important to explain the observed correlations without elaborate 
computer models. The aim 
of this paper is to explain the  correlations, by taking into account the non-linear mixing of modes through quadratic order  in ``single-shot" hydrodynamics \cite{Gardim:2011xv,Qiu:2011iv}. 

\section{Methodology}
\label{sec2}

\begin{figure}[htbp]
\begin{center}
\includegraphics[width=0.49\textwidth]{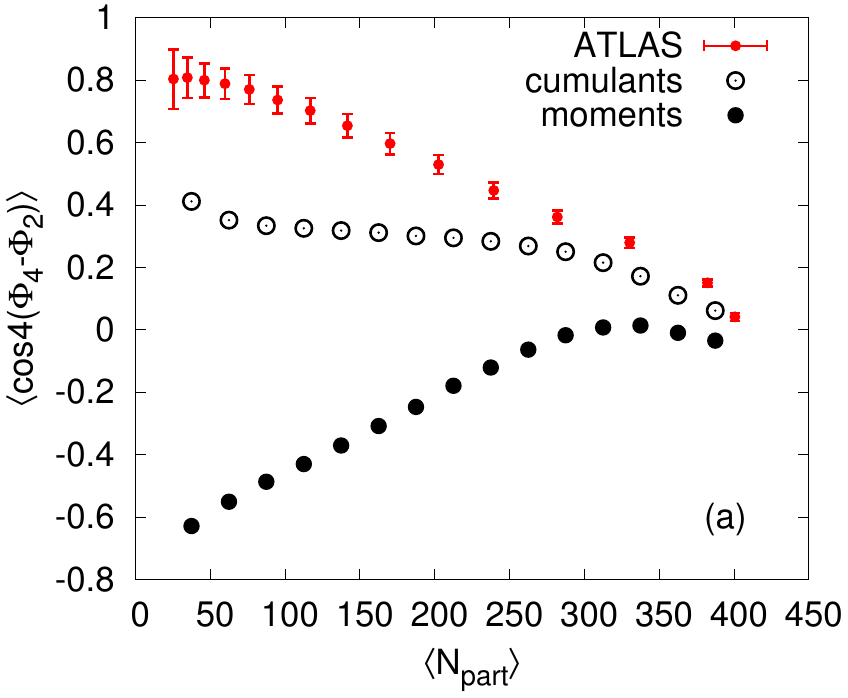}
\includegraphics[width=0.49\textwidth]{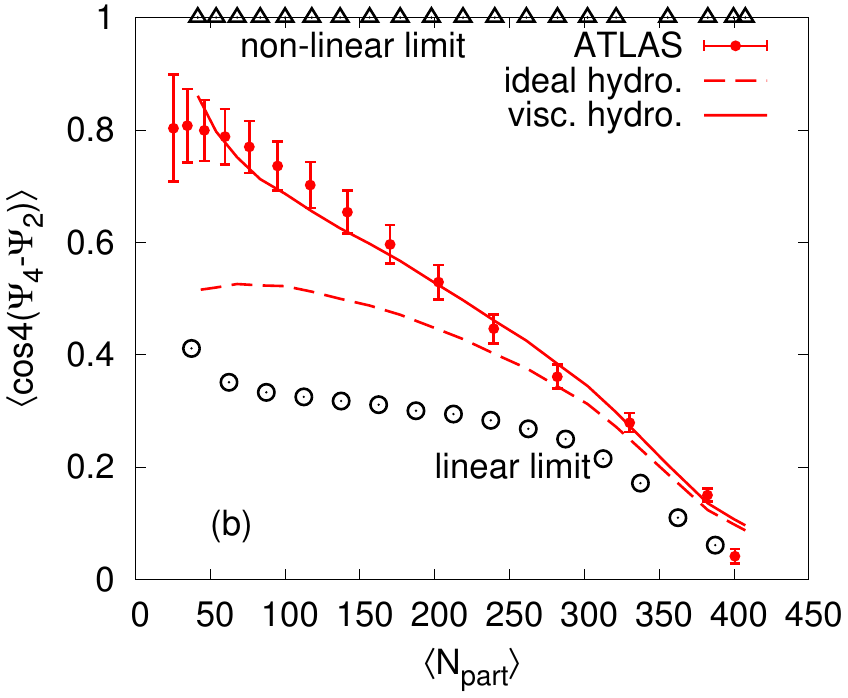}
\end{center}
\caption{(a) Initial participant plane $\Phi_2$ and $\Phi_4$ correlation, as a function of centrality. 
Simulations are done with PHOBOS Glauber model with $\Phi_4$ from both cumulants and moments definitions.
The data points are extracted from ATLAS \cite{Jia:2012sa} for comparison. (b) Event plane $\Psi_2$ and
$\Psi_2$ correlation, with predictions made by ideal and viscous hydrodynamics.}
\label{fig_ini}
\end{figure} 

Event plane correlations can be obtained in single-shot hydrodynamics using
a quadratic response formalism \cite{Teaney:2012ke}. As  detailed in \cite{Teaney:2012ke},  the  response formalism
uses  the cumulant expansion (as opposed to a moment expansion) to 
characterize the initial geometry \cite{Teaney:2010vd}. The
medium response to a given cumulant is calculated at linear 
and quadratic order by deforming a Gaussian distribution and 
calculating the response coefficients with  ideal and viscous hydrodynamics.

%The cumulant based definitions avoids double
%counting  the non-linearity of flow response.  
For example, the fourth order cumulant
$C_4$ and angle $\Phi_4$  are closely related to  the fourth order eccentricity 
$\llangle r^4 e^{i4\phi_s} \rrangle$, and  characterize the quadrangular deformation  of energy density in the
transverse plane
\[
   C_4 e^{i4\Phi_4} = \llangle r^4 e^{i4\phi_s} \rrangle  - 3 \llangle r^2 e^{i2\phi_s } \rrangle^2 \, .
\]
Using cumulants rather than moments to characterize the initial
state geometry leads to different correlations between the 
angles $\Phi_n$. \Fig{fig_ini}a shows the correlations 
between $\Phi_2$ and $\Phi_4$ with cumulant and moment based
definitions.

The particle spectrum is expanded in harmonics
\be
\label{dn_spec}
\frac{dN}{d\phi_p}=\frac{N}{2\pi}\left[1+\sum_n v_ne^{in(\phi_p-\Psi_n)}+c.c.\right]
=\frac{N}{2\pi}\left[1+\sum_n z_ne^{in\phi_p}+c.c.\right] \, ,
\ee
and each harmonic and angle, $z_n \equiv v_n e^{-in\Psi_n}$, will receive
contributions from both the linear and quadratic response to a given cumulant. 
For $v_4$ for example,  the dominant quadratic response is proportional to $(\epsilon_2e^{i2\Phi_2})^2$,
and the linear response is proportional to the fourth order cumulant $C_4e^{i4\Phi_4}$
\be
\label{z4}
z_4=v_4e^{-i4\Psi_4}=w_4e^{-i4\Phi_4}+w_{4(22)}e^{-i4\Phi_2} \, , 
\ee 
where $w_4$ and $w_{4(22)}$  indicate linear response to $C_4$ and the non-linear  response to $\epsilon_2^2$.

\begin{figure}[htbp]
\begin{center}
\includegraphics[width=1\textwidth]{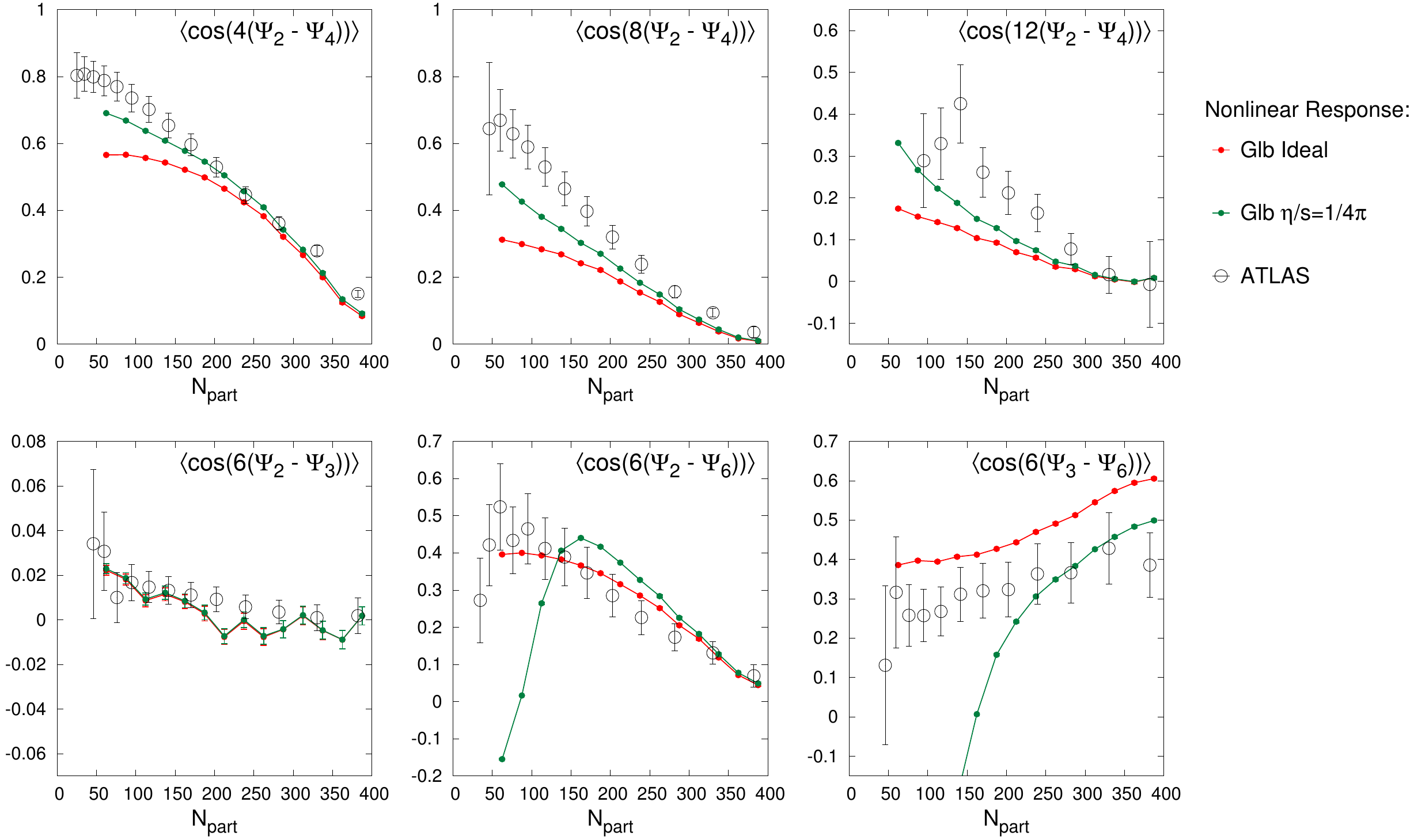}
\end{center}
\caption{Two-plane correlations of event plane correlations measured at ATLAS\cite{Jia:2012sa}. }
\label{fig_2plane}
\end{figure}

\begin{figure}[htbp]
\begin{center}
\includegraphics[width=1\textwidth]{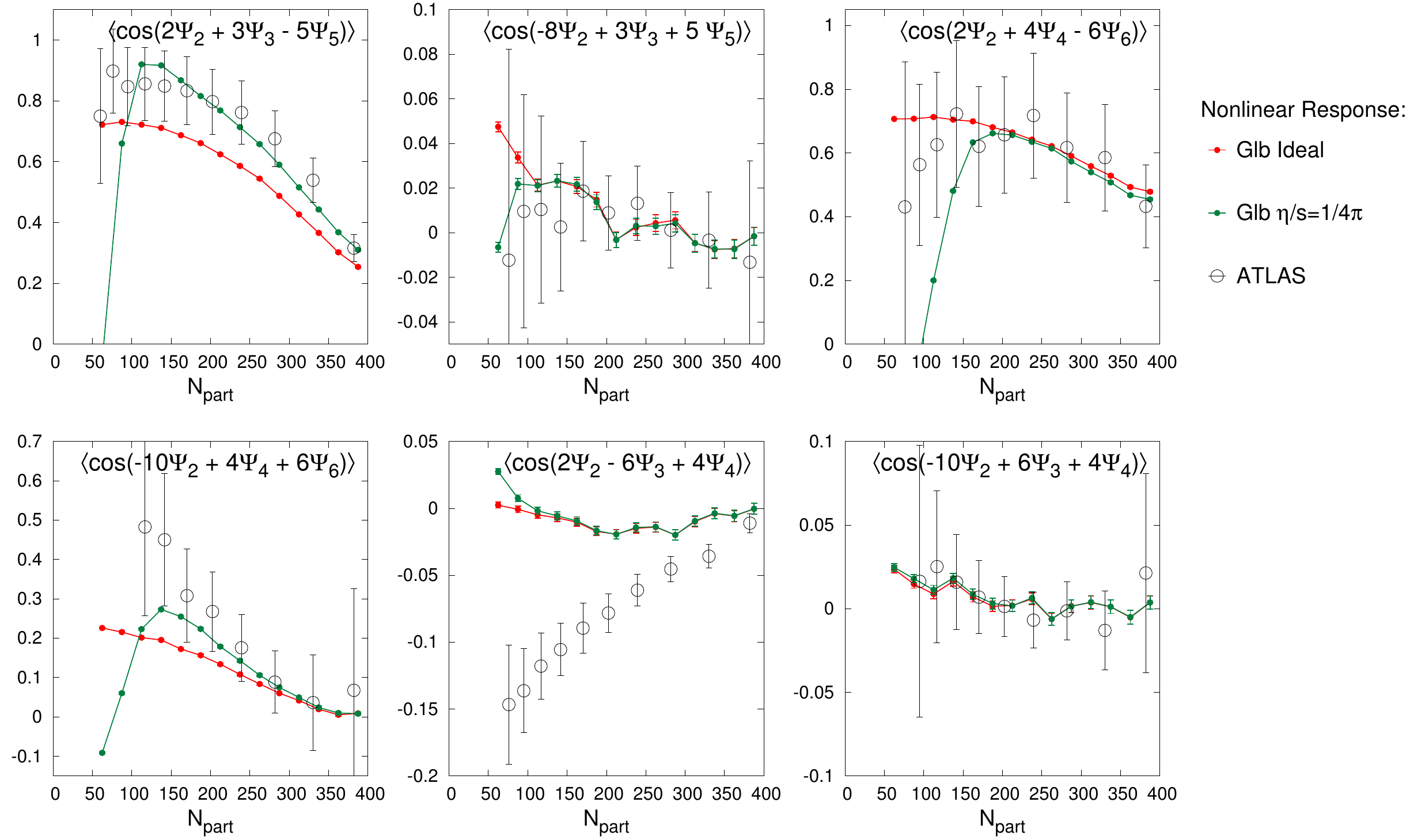}
\end{center}
\caption{Three-plane correlations of event plane correlations measured at ATLAS\cite{Jia:2012sa}. The non-linear
flow generations we considered in these calculations are: $w_{3(12)}$ for $v_3$, $w_{4(22)}$ for $v_4$,
$w_{5(23)}$ for $v_5$, and $w_{6(33)}$, $w_{6(24}$, $w_{6(222)}$ for $v_6$. Linear flow generation $w_6$ is
ignored because of its comparably small contributions. }
\label{fig_3plane}
\end{figure}

It is straightforward then to express the event plane correlations using the complex form of harmonic flow $z_n$
in \eq{dn_spec}, {\it e.g.} 
\be
\label{42cor_form}
\cos4(\Psi_4-\Psi_2)=\frac{\mbox{Re}[z_2^2z_4^*]}{|z_2|^2|z_4|}
=\frac{\cos4(\Phi_4-\Phi_2)w_4+w_{4(22)}}{|w_4e^{-i4\Phi_4}+w_{4(22)}e^{-i4\Phi_2}|}.
\ee  
\eq{42cor_form} shows the typical structure of event plane correlations 
in the framework of non-linear response.
In the numerator and denominator  one term is determined 
by the linear response, $w_4$, and one term  is determined from quadratic response, $w_{4(22)}$. In more involved cases, 
where higher order non-linear corrections are also included, there
are interference terms between the linear and 
quadratic response. The combined effects of all these linear, non-linear, and interference terms
determines the behavior of the event plane correlations.  

\fig{fig_ini}(b) shows that
if linear response $w_4$ dominates the $v_4$ signal, the event plane correlation
reduces to its corresponding initial participant plane correlation, $\llangle
\cos4(\Phi_4-\Phi_2) \rrangle$. On the other hand, if the $v_4$ signal
is dominated by the non-linear response, $w_{4(22)}$, the event planes
$\Psi_4$ and $\Psi_2$ would be perfectly correlated. Referring to these two
limits as the linear and non-linear limits respectively,
the observed event plane correlations can be understood as an 
average of these two limits, with the actual value controlled by the relative magnitudes of the two response coefficients, $w_4$ and $w_{4(22)}$.  
When these response coefficients are computed with ideal and viscous hydrodynamics and averaged as in \eq{42cor_form}, the event plane correlations  seen 
in \Fig{fig_ini}(b) are quantitatively reproduced by the response formalism. A more complete comparison of the response formalism to the two and
three plane correlations is given in \Fig{fig_2plane} and \Fig{fig_3plane}.
 
The non-linear flow response dominates correlation for non-central collisions
and for larger values of $\eta/s$. Thus, we find that the observed
non-trivial event plane correlations become stronger toward non-central
collisions as seen in \fig{fig_2plane} and \fig{fig_3plane}. Also as the shear
viscosity is increased, the predicted correlations increase and approach  the
non-linear limit seen in \Fig{fig_ini}(b). The observed correlation
patterns involving $\Phi_6$ have a rich structure. In the response formalism
this rich pattern of correlations is due to the fac that two quadratric response coefficients,
$w_{6(24)}$ and $w_{6(33)}$, contribute to $v_6$. 

\section{Summary and discussion}
\label{sec5}

Using the PHOBOS MC-Glauber model, $\eta/s=1/4\pi$, and a freeze-out temperature
$T_{\mbox{fo}}=150\,{\rm MeV}$, the non-linear response formalism  qualitatively reproduces the observed two-plane and three-plane correlations, with 
one exception,  $\dlangle\cos(2\Psi_2-6\Psi_3 + 4\Psi_4) \drangle$. This correlator is currently under investigation.
Since we have included only the quadratic response to the lowest deformations,
this level of agreement is acceptable. In addition, a preliminary comparison of our predictions to event-by-event hydrodynamics
%and for the calculations which aim mainly at interpretations\footnote{Many of the parameters used here 
%are selected for general discussions, thus are not fine tuned with respect to experiment.}, 
also shows qualitative agreement \cite{Qiu:2011iv}. In particular,  the response formalism reproduces the  shear viscosity dependence of the correlations found in event-by-event hydrodynamics.  A more complete comparison to event-by-event hydrodynamics and the data is reserved for future work.

\section*{Acknowledgements}
We would like to thank U. Heinz and Z. Qiu for the helpful discussions.
This work is supported in part by the Sloan Foundation and by the Department of Energy 
through the Outstand Junior Investigator program DE-FG-02-08ER4154.

\section*{References}

%\begin{thebibliography}{00} 
\bibliography{iopart-num}

%\end{thebibliography}
\end{document}